\documentclass{elsart}

\usepackage{graphicx}

\def\plotone#1{\centering \leavevmode
 \includegraphics[width=.95\columnwidth]{#1}}

\begin{document}
\begin{frontmatter}
\title{The Infrared Sky}
\author{E. L, Wright}
\address{UCLA Astronomy, PO Box 951562, Los Angeles, CA 90095-1562, USA}

\begin{abstract}
The infrared sky from space is the sum of a cosmic signal from
galaxies, quasars, and perhaps more exotic sources; and foregrounds
from the Milky Way and from the Solar System.  At a distance of 1 AU
from the Sun, the foreground from interplanetary dust is very bright
between 5 and 100 $\mu$m, but ``very bright'' is still several million
times fainter than the background produced by ground-based telescopes.
In the near infrared 1-2.2 $\mu$m range the space infrared sky is a
thousand times fainter than the OH nightglow from the Earth's
atmosphere. As a result of these advantages, wide-field imaging from
space in the infrared can be an incredibly sensitive method to study
the Universe.
\end{abstract}

\end{frontmatter}

\begin{figure}[t]
\plotone{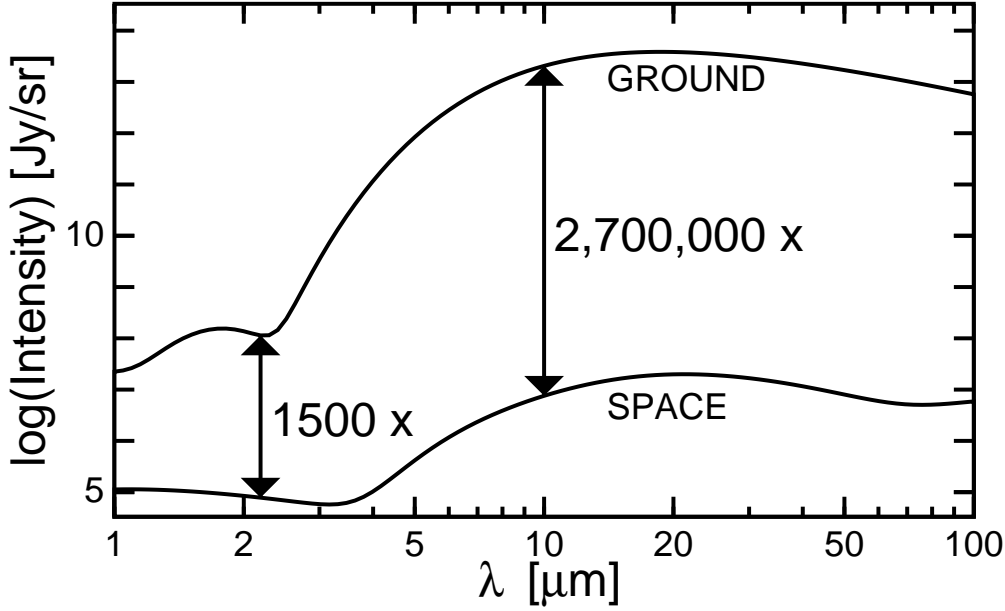}
\caption{The infrared sky brightness from the Earth, dominated by the OH nightglow
and thermal emission from ambient temperature telescopes at $0^\circ$~C, compared
to the sky brightness from space at 1 AU from the Sun, dominated by scattered
Sunlight and thermal radiation from interplanetary dust.
\label{fig:SvsG}}
\end{figure}

\begin{figure}[t]
\plotone{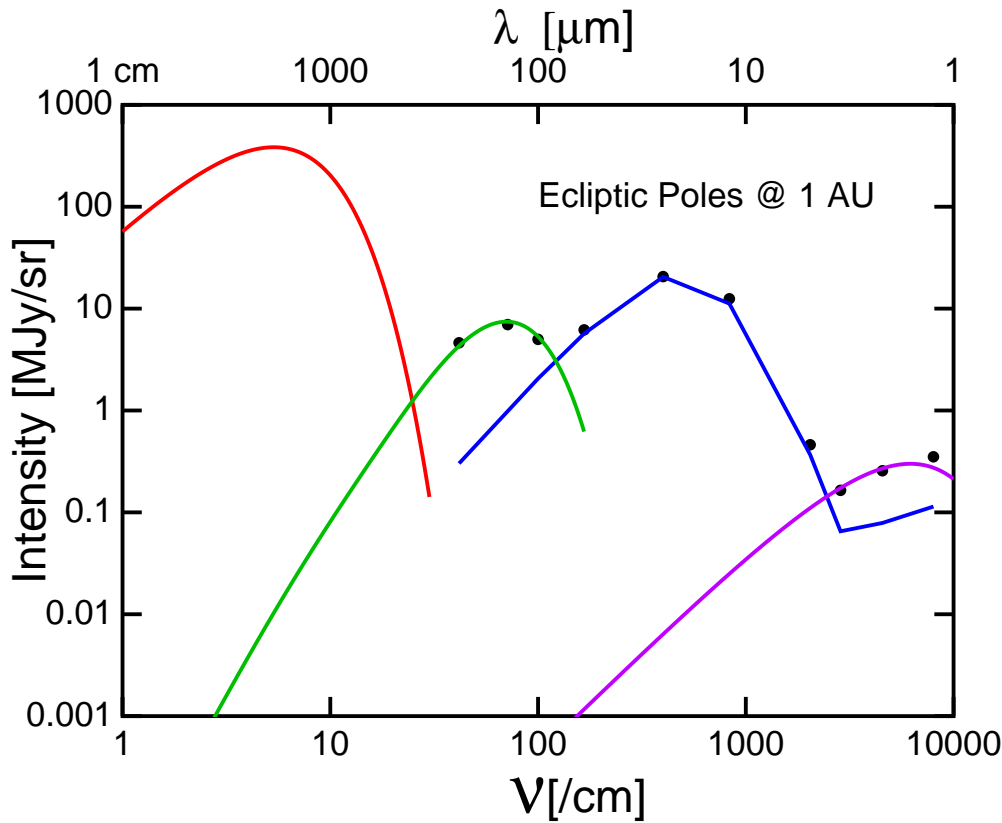}
\caption{The intensity averaged over the DIRBE mission of the ecliptic poles.
The dots are the DIRBE data, while the red curve is the FIRAS data and a
Planck function at 2.725 K, the green curve is a scaled version of the
FIRAS spectrum of the interstellar dust in the Milky Way, the blue line
is a model of the zodiacal dust fit to the DIRBE data, and the magenta
curve is an approximation to galactic starlight.
\label{fig:Inu}}
\end{figure}

\begin{figure}[t]
\plotone{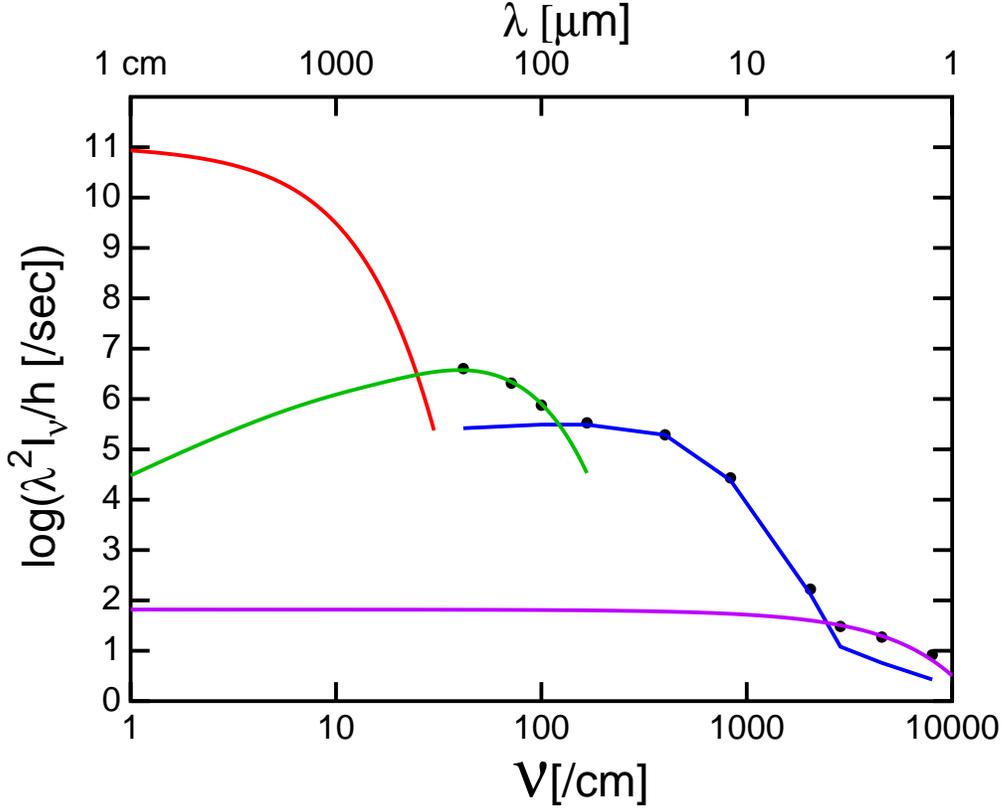}
\caption{The photon rate on a diffraction-limited pixel ($A\Omega = \lambda^2$)
in a band with $\Delta\lambda = \lambda$.  To achieve background-limited
performance, the dark current should be less than 1\% of these values.
\label{fig:DL-photons}}
\end{figure}

\begin{figure}[t]
\plotone{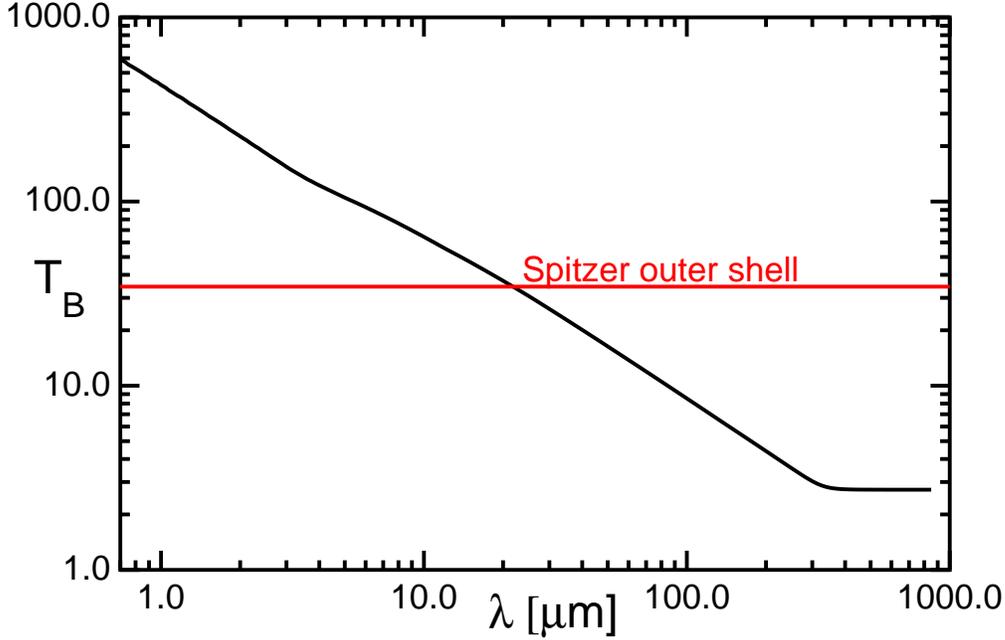}
\caption{The Planck brightness temperature of the infrared sky in space 1 AU
from the Sun.  The temperature achieved by passively cooling the Spitzer
Space Telescope outer shell is indicated.\label{fig:TB}}
\end{figure}

\begin{figure}[t]
\plotone{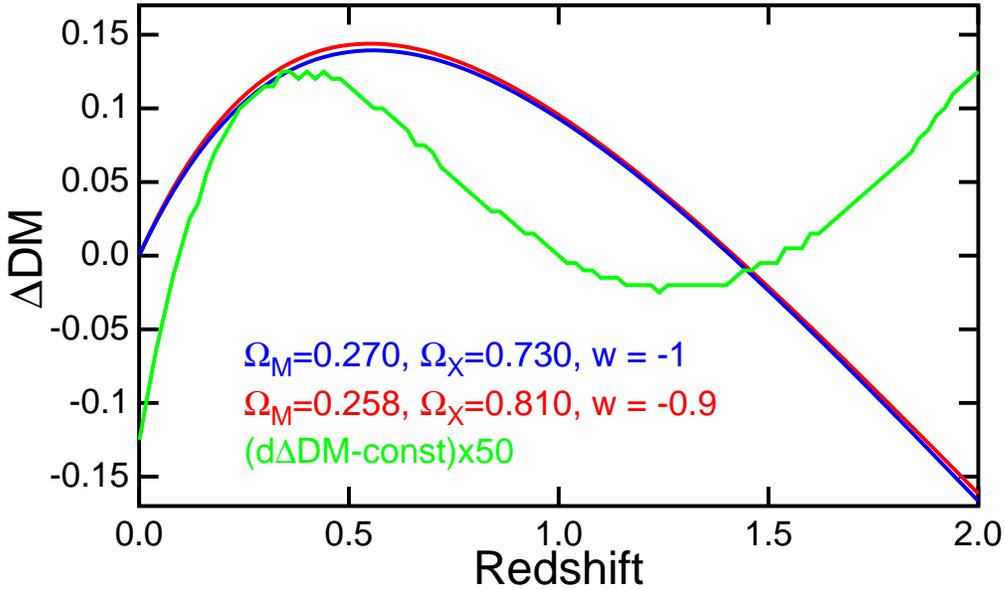}
\caption{Degeneracy between the WMAP concordance model with $w = -1$ and a
model with $w = -0.9$.  The difference is only $\pm2$ millimag.
Supernovae alone cannot determine $w$.
\label{fig:dw0}}
\end{figure}

\begin{figure}[t]
\plotone{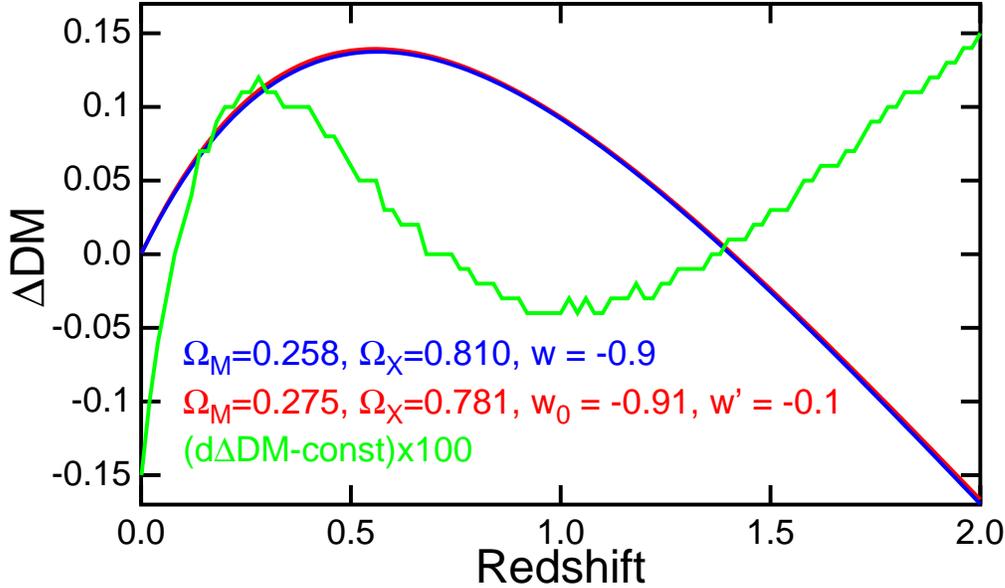}
\caption{Degeneracy between the model in Figure \protect\ref{fig:dw0}
with $w^\prime = 0$ and the a model with $w^\prime = 0.1$. The
difference is only $\pm1$ millimag.
Supernovae alone cannot determine $w^\prime$.
\label{fig:dw1}}
\end{figure}

\begin{figure}[t]
\plotone{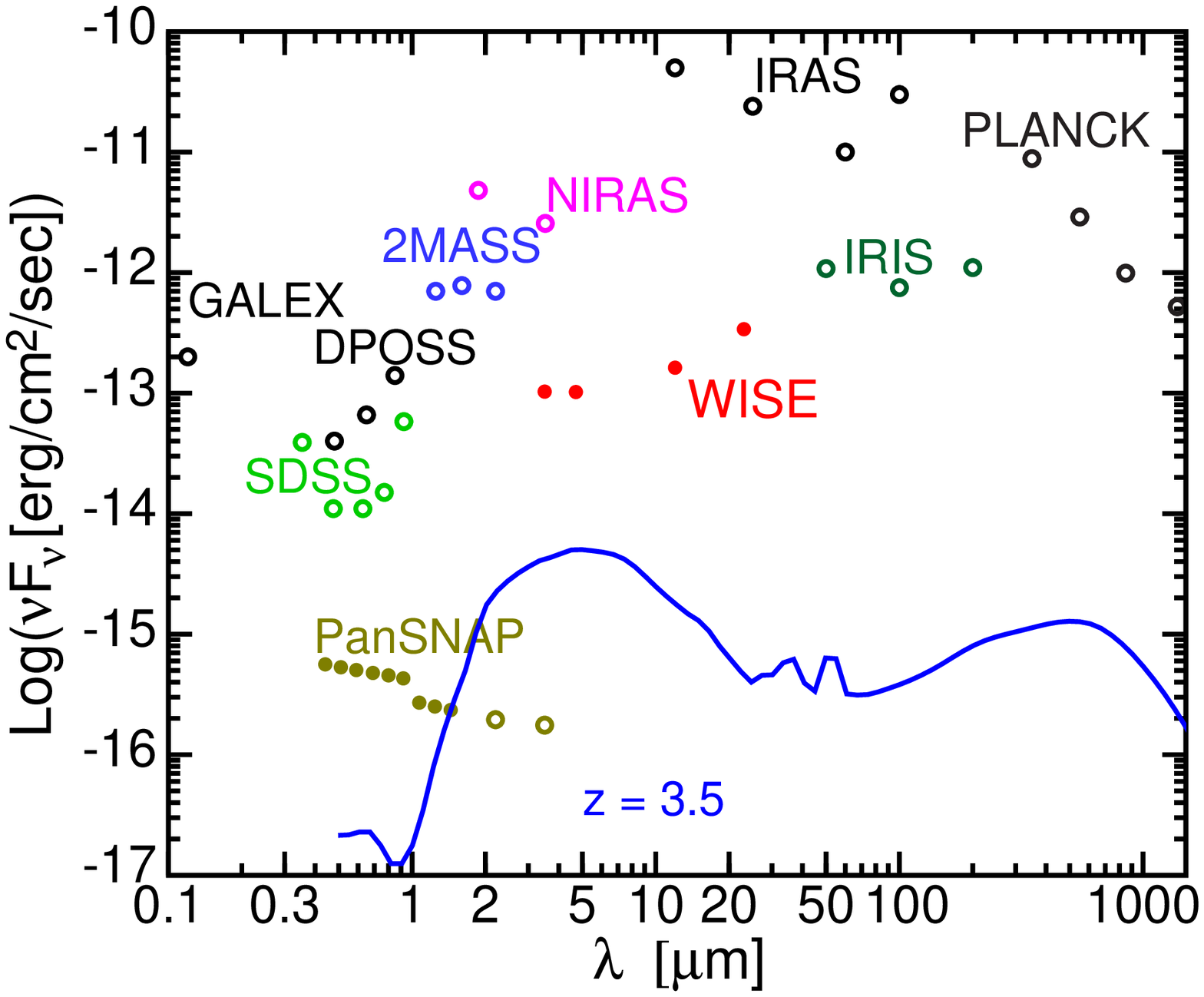}
\caption{$5\sigma$ point source sensitivities of various existing or
planned wide-field surveys.  NIRAS is the 1988 SMEX proposal
by Giovanni Fazio, E. L. Wright and others that led indirectly to the 2MASS
survey.  The Wide-field Infrared Survey Explorer is a MidEx under study
that has a planned launch in 2008.  The panoramic survey that might be performed
by SNAP is labeled PanSNAP.  The open circles at 2.2 and 3.5 $\mu$m are
not currently planned for SNAP, but would be 20 times more sensitive to
old stellar populations in $z = 3$ galaxies than longest currently
planned band in SNAP.
\label{fig:prev}
}
\end{figure}

\section{Introduction}

The infrared sky on Earth is very bright, leading George Rieke to compare
infrared astronomy to observing stars in the daytime with a telescope made out 
of fluorescent lights.  
From space the situation is much improved, with the
total sky brightness down by three orders of magnitude in the near infrared
and more than six orders of magnitude in the thermal infrared.  
Figure \ref{fig:SvsG} shows the sky brightness on the Earth compared to the
sky brightness from space 1 AU from the Sun.

\section{The Total Infrared Sky}

The Diffuse
Infrared Background experiment (DIRBE) on the Cosmic Background Explorer made
observations of the total intensity of the sky in ten bands from 1.25 to
240 $\mu$m \cite{DIRBE}.  
The total intensity at the ecliptic poles from DIRBE and FIRAS
data is shown in Figure \ref{fig:Inu}.  One can clearly see two
windows in this foreground emission:  one at 200-500 $\mu$m and
another at 3 $\mu$m.  These windows provide an opportunity for
extremely sensitive imaging from space.  The curves plotted on this
graph are the CMB, a 2.725~K blackbody, a scaled version of the
FIRAS spectrum of the Milky Way \cite{FIRASMW}, 
\begin{equation}
I_\nu = 1.54 \times 10^{-4} \left(\frac{100\;\mu\mbox{m}}{\lambda}\right)^2
\left(B_\nu(20.4)+6.7B_\nu(4.77)\right)
\end{equation}
a model for the interplanetary dust \cite{DIRBEPS}, 
and an estimate for the stellar
light in the Milky Way: $5 \times 10^{-13} B_\nu(10^{3.5}\;\mbox{K})$.

\section{Telescope Cooling Requirements}

The brightness of the infrared sky defines the maximum temperature a
telescope can have without degrading the mission sensitivity.  Since
optical surfaces normally have rather low emissivities, cooling the optics
to the Planck brightness temperature of the infrared sky will lead to
minimal degradation of performance.  Figure \ref{fig:TB} shows this
temperature as function of wavelength.  Also shown on this plot is the
temperature of the outer shell of the Spitzer Space Telescope, which is
passively cooled in an environment 1 AU from the Sun but far enough
away from the Earth that it is not a significant thermal source.
Clearly it is rather easy to use a passively cooled telescope in the
3 $\mu$m window in the zodiacal foreground.

\section{Detector Performance Requirements}

When designing space-based wide-field imaging missions one naturally wants to
achieve background-limited sensitivity.  Figure \ref{fig:DL-photons}
show the same data as Figure \ref{fig:Inu} but in a different representation.
Here the y-axis shows the photon rate on a diffraction-limited pixel
with 100\% transmission and 100\% bandwidth.  Furthermore one usually
oversamples the diffraction-limited point spread function so a pixel in
a real experiment will receive only about a few percent of this photon rate.  Clearly
the dark current requirements for background-limited imaging are well below
1 e/sec in the 3 $\mu$m window in the zodiacal light, but 10,000 e/sec would
be acceptable at 300 $\mu$m.

A simple order of magnitude estimate shows that the detectors must be much
colder than the optics temperature requirement.  The photon arrival rate on a
unit area in a blackbody is $O(\nu^2 e^{-x}/c^2)$ where 
$x = h\nu/kT_B$ is $>> 1$.  In fact $x$ is usually 25-40 for the space-based
backgrounds in Figure \ref{fig:TB}.  On the other side of the detector
surface one has phonons that can also be detected, and the phonon arrival
rate is $O(\nu^2 e^{-x^\prime}/c_s^2)$ where $c_s$ is the sound speed
and $x^\prime = h\nu_{min}/kT_D$.  Here $h\nu_{min}$ is the minimum energy
that can produce a charge carrier in the detector, so clearly $\nu_{min}$
has to be less than the observing frequency $\nu$.  Usually $\nu_{min}$
corresponds to the cutoff wavelength of the detector but in some cases it
can be smaller.  Because $c^2/c_s^2 \approx 10^{10}$, one needs to have
$x^\prime > x+23$, and thus the detectors usually need to be 2-3 times
colder than the optics.  Even so, there is no problem running passively
cooled optics and HgCdTe detectors at full natural background-limited
sensitivity in the 3 $\mu$m window through the zodiacal light.

\section{JDEM Requirements}

The Joint Dark Energy Mission (JDEM) is supposed to provide data about
the dark energy.  One convenient but not physically motivated
parametrization of the dark energy is in terms of a time varying ratio
between the pressure and the density, with $w \equiv P/\rho c^2 =
w_\circ + 2 w^\prime (1-a)$ where $a(t) \equiv 1/(1+z)$ is the scale
factor of the Universe \cite{SNAP}.  When $w \neq -1$ there should be a
dynamical coupling between the dark energy and matter, leading to an
inhomogeneous dark energy density, and this is neglected in the simple
$w$, $w^\prime$ model.  But the luminosity distance can easily be
computed in the $w$, $w^\prime$ model allowing for quick parameter
sensitivity studies.  It turns out that the luminosity distance {\it
vs.} redshift data for supernovae only constrain two directions in the
4 parameter space spanned by $\Omega_M$, $\Omega_X$, $w$ and
$w^\prime$.  Figure \ref{fig:dw0} and Figure \ref{fig:dw1} show that
the other two directions are essentially unconstrained by observations
of supernova brightness in the redshift range $0 < z < 2$.

All of the calculations that claim that supernova data can determine
$w$ and $w^\prime$ have made some poorly justified assumptions.
For example, the assumption that the Universe is flat
($\Omega_M + \Omega_X = 1$) is often made.  But the flatness of the
Universe is clearly something that needs to be determined from the
data.  Often the assumption of a flat Universe is justified on the
basis of simplicity, but simplicity also implies that $\Omega_X = 0$.
Or the CMB is cited in justifying the assumption of a flat Universe.
But the CMB data does not say the Universe is flat!  The CMB data are
perfectly happy with a closed $\Omega_M = 1.3$, $\Omega_X = 0$ model.
Only when combined with the supernova data, and the assumption that
$w = -1$, do the CMB data imply a flat Universe.  To then turn this
around and try to find $w$ is clearly circular reasoning.

Where then can the additional constraints on $\Omega_M$, $\Omega_X$,
$w$ and $w^\prime$ be found?  The CMB data, in particular the location
of the acoustic peaks, provide one independent constraint.  And weak
lensing surveys offer another constraint by measuring the growth of
perturbations as a function of redshift.  But weak lensing observations
require a very high density of background sources, and thus a high
sensitivity to typical galaxies.  Typical stars in typical galaxies are
much redder than supernovae, and galaxies are redshifted as well,
leading to a requirement for longer wavelength observations.  Figure
\ref{fig:prev} shows how the sensitivity of a proposed panoramic survey
by SNAP to $z=3$ background galaxies could be increased by a factor of
20 by extending the wavelength coverage out to 3.5 $\mu$m.  Note that I
recalculated the plotted sensitivities for all the SNAP bands and
generally agree with published estimates \cite{SNAP}.

This sensitivity at 3.5 $\mu$m is two magnitudes fainter than the
faintest reported number counts from Spitzer \cite{IRACN}.  Since the
Spitzer counts exceed $10^5$ per square degree, the counts from a
panoramic SNAP survey would be about $10^{5.4}$ per square degree if
the wavelength coverage of SNAP is extended to 3.5 $\mu$m.

But this sensitivity at 3.5 $\mu$m will require rethinking
the implementation of JDEM.  The Hubble Space Telescope was
designed in the 1980's, without planning for sensitive infrared
instruments, and as a result has heated mirrors made out of
ULE glass.  ULE is not ultra-low expansion except in a narrow
range of temperature near room temperature.  The recently
released Hubble Ultra-Deep Field shows the unfortunate
consequences of this decades old decision:  even though the
NICMOS exposure time per position is only a few percent of the
ACS exposure, the NICMOS image clearly goes to higher redshift.
And NICMOS cannot even use its 2.2 $\mu$m capability because of
the heated mirrors.

The James Webb Space Telescope \cite{JWST}
will use passively cooled optics and
actively cooled detectors to operate in the $0.6 < \lambda < 25\;\mu$m
range with high sensitivity.  The Wide-field Infrared Survey Explorer
(WISE) will survey the entire sky with sensitivities better than the
requirements shown on Figure \ref{fig:prev}, and thus provide a sky
survey suitable for planning JWST observations.

WISE \cite{WISE}
will have 4 bands at 3.5, 4.7, 12 \& 23 $\mu$m and will be able to
detect old, cold brown dwarfs in the Solar neighborhood as well as
Ultraluminous Infrared Galaxies at redshifts up to $z=3$.  WISE will be
in a nearly polar Sun-synchronous low Earth orbit like IRAS and COBE
and will scan a circle perpendicular to the Earth-Sun line once per
orbit while looking away from the Earth.  After 6 months the precession
of the orbit will have swept this scan circle across the entire sky.

WISE will take exposures every 11 seconds with $1024\times1024$ pixel 
arrays in each of its 4 bands.  The 40 cm diameter WISE telescope will
scan at a constant rate while an internal scan mirror freezes an
image of the sky on the arrays.  This scan method is very similar to
the method employed by the 2MASS survey.

The current catalogs of stars close to the Sun are surprisingly incomplete.
In fact, the third closest star to the Sun was discovered in 2003 with a
proper motion of $5^{\prime\prime}$ per year \cite{Teegarden}!
When brown dwarfs are included, it is likely that only one-third of the stars
within 5 pc have been identified.  Most of the missing objects will be
old low mass brown dwarfs with luminosities considerably smaller
than the $2.6 \times 10^{-6}\;\mbox{L$_\odot$}$ of 2MASS 1415-09 \cite{Vrba}.  
WISE will be
able to detect brown dwarfs down to $10^{-8}\;\mbox{L$_\odot$}$ at the
distance of Proxima Cen, and there could be two such objects within the
10 pc$^3$ closer to the Sun than Proxima Cen.

At much greater distances our knowledge of active galactic nuclei, QSOs
and ultra-luminous starburst galaxies is also incomplete because these 
objects are often obscured by tremendous amounts of dust.  WISE should be
able to detect about $10^7$ galaxies at 23 $\mu$m and 5-6\% of these will
have redshifts $z > 2$.

All of these objects that can be found by WISE will be of great interest
for followup observations using the JWST.  A 2 meter diameter, passively cooled,
wide field telescope working in the near infrared would fill a valuable
niche between the wide but shallow survey of WISE and the ultra-sensitive
but narrow field capability of JWST.

\section{Conclusion}

The Joint Dark Energy Mission goals require many different
kinds of data, not just supernova brightness {\it vs.} redshift.
A high sensitivity to faint redshifted galaxies will be essential,
and in the space environment 1 AU from the Sun the highest
sensitivity is obtained at 3 $\mu$m \cite{ELWST}.  Thus JDEM should 
cover the wavelength range $\lambda < 4\;\mu$m, and this can easily
be achieved using passively cooled optics and detectors.

\end{document}